\documentclass[a4paper,conference]{IEEEtran}
\IEEEoverridecommandlockouts
\usepackage{cite}
\usepackage{amsmath,amssymb,amsfonts}
\usepackage{algorithmic}
\usepackage{cancel}
\usepackage{graphicx}
\usepackage{textcomp}
\usepackage{xcolor}
\usepackage{amsmath}
\usepackage{placeins}
\usepackage{longtable}
\def\BibTeX{{\rm B\kern-.05em{\sc i\kern-.025em b}\kern-.08em
    T\kern-.1667em\lower.7ex\hbox{E}\kern-.125emX}}
\begin{document}
\title{Linear approximation of CPM signals for a reduced-complexity, multi-mode telemetry transmitter\\
}

\author{\IEEEauthorblockN{Francesco Silino}
\IEEEauthorblockA{\textit{University of Pavia, Italy} \\
francesco.silino01@universitadipavia.it}
\and
\IEEEauthorblockN{Fabio Dell'Acqua}
\IEEEauthorblockA{\textit{University of Pavia} \\
\textit{\& CNIT Consorzio Nazionale}\\
Interuniversitario per le \\
Telecomunicazioni - Unità di Pavia\\
Pavia, Italy\\
fabio.dellacqua@unipv.it}
\and
\IEEEauthorblockN{Pietro Savazzi}
\IEEEauthorblockA{\textit{University of Pavia} \\
\textit{\& CNIT Consorzio Nazionale}\\
Interuniversitario per le \\
Telecomunicazioni - Unità di Pavia\\
Pavia, Italy\\
pietro.savazzi@unipv.it}
\and
\IEEEauthorblockN{Anna Vizziello}
\IEEEauthorblockA{\textit{University of Pavia} \\
\textit{\& CNIT Consorzio Nazionale}\\
Interuniversitario per le \\
Telecomunicazioni - Unità di Pavia\\
Pavia, Italy\\
anna.vizziello@unipv.it}
\and
\IEEEauthorblockN{Diego Biz}
\IEEEauthorblockA{\textit{Temis Srl} \\
\textit{Corbetta (MI), Italy}\\
diego.biz@temissrl.com}
\and
\IEEEauthorblockN{Federico Brega}
\IEEEauthorblockA{\textit{Temis Srl} \\
\textit{Corbetta (MI), Italy}\\
federico.brega@temissrl.com}
}
\markboth{Journal of \LaTeX\ Class Files,~Vol.~14, No.~8, August~2015}%
{Shell \MakeLowercase{\textit{et al.}}: Bare Demo of IEEEtran.cls for IEEE Journals}
%



\maketitle

\begin{abstract}
In space applications, hardware (HW) implementation is made more expensive not only by the levels of performance required, but also by complex and rigorous HW qualification tests. Reducing qualification cost and time is thus a key design requirement.
In this paper, a new versatile transmitter is proposed for space telemetry, capable of soft-switching across different linear and continuous phase modulation schemes while maintaining the same hardware structure.
This permits a single HW qualification  to ``cover'' diverse uses of the same hardware, and thus avoid re-qualification in case of configuration changes.
The envisaged solution foresees the use of a single filter, suitable not only for linear modulations such as M-QAM, but also for continuous phase modulation methods. At this stage, we focus on pulse code modulation/frequency modulation (PCM/FM), for which we propose a minimum mean square error (MMSE) algorithm. The proposed algorithm, which adds to the system flexibility and effectiveness, may use a single first filter based on Laurent decomposition for initialization, if needed.
Performances are assessed using the mean square error (MSE) measure between the proposed MMSE-modulated signal and the completely modulated signal. Simulation results confirm that the proposed algorithm leads to MSE values that are lower than the case of Laurent decomposition using the first component only. 
\end{abstract}


%
\IEEEpeerreviewmaketitle

\section{Introduction}

%
%
%
%
In space telemetry, which consists of the transmission of data from a satellite to its ground station \cite{9581544}, pulse code modulation/frequency modulated (PCM/FM) and shifted offset quadrature phase shift keying (SO-QPSK) are the most commonly used modulation schemes \cite{9044986,9530180}. Their success is due to their low speed and high robustness, which allow spanning the long distances from the base station to the space platform and vice versa. On the other hand, these schemes achieve low rates in relation to band occupation, particulary so for PCM/FM. New techniques are needed to improve both flexibility and performance of the transmitters employing these modulations. 

In this work, the focus is on improving the Laurent decomposition of PCM/FM modulation, by introducing a new type of approximation of the PCM/FM signal.  
Laurent decomposition consists of approximating a continuous-phase modulated (CPM) signal with a superposition of pulse amplitude modulated (PAM) components. The non-linearity is moved upstream with respect to the filter. Summing up components after filtering allows reconstructing the original signal. This technique reduces the complexity of the modulation, if the first component only is used to approximate the PCM/FM \cite{artlaur,pamdec}.
Laurent decomposition is usually employed at the receiver side to justify a linear approximation of a generic PCM signal, for estimating the unknown modulation index~\cite{zhipeng_h} or channel variations~\cite{zhramichanest,jsac98}.

At the transmitter side, the main drawback is that a perfect reproduction of the original PCM/FM signal requires two filters and two streams of pseudosymbols (computed form the original bitstream), if $L\ge 2$ is used. Increasing $L$, i.e. the phase pulse duration in symbol intervals, results in inflating the number of filters. It is also known that in Laurent decomposition most of the energy is contained in the first filter signal component (more than 90\% for PCM/FM) and in the first two components for SO-QPSK or similar modulations.

The goal of this work is to introduce a new type of decomposition that uses the same pseudo-symbols as Laurent decomposition related to the first filter $c_0$, but introducing a new filter computed using the Wiener-Hopf equations for minimizing the MSE. This allows improving the approximation performance, using a new filter that unexpectedly has a smaller number of taps with respect to $c_0$. This means using only one component and reducing the complexity of devices.

The main advantage of the proposed approach is related to the usage of only one linear filter for partial response modulations, for which the Laurent decomposition requires multiple linear filters.

This new technique is here used at the transmitter side to allow the usage of different linear and nonlinear modulations schemes, by changing only a few settings related to the transmit filter and the way bits are mapped into modulated symbols. Indeed, the proposed transmitter is able to quickly adapt modulation from PCM/FM to a different one, also linear. The ongoing work is devoted to extending the proposed approach to SO-QPSK modulations, by leveraging the decomposition in \cite{PerrinRice}, and to include M-QAM modulation schemes in the reconfigurable transmitter \cite{LamoralCoines2021CCSDS1T}.

To summarize, the main novelties presented in this paper are the following:
\begin{itemize}
    \item a new type of approximated decomposition, based on the same pseudo-symbols used as input of the first filter that results from the Laurent approach;
    \item the experimental implementation of the proposed scheme in a space telemetry transmitter.
    \item a new transmitter scheme, capable of switching across CPM modulation techniques without changing the internal - space-qualified - structure of the device, by leveraging a Software-Defined-Radio (SDR) approach.
\end{itemize}
    
The rest of this paper is organized in four sections. Section II describes the PCM/FM modulation and recalls the Laurent decomposition technique. Section III details the proposed new implementation methodology, whereas in section IV results and performance of the different techniques are discussed, including comparison with field programmable gate array (FPGA)-based results. Section V  presents the conclusions and future lines of research.

 \section{PCM/FM modulation}
 
 \subsection{CPM signal}
 
A CPM signal can be expressed as:
 
\begin{equation}
\label{eq:s}
  s(t,{\alpha})= e^{j\psi(t,{\alpha})}   
\end{equation}

where:

\begin{equation}
  \psi(t,{\alpha})=2h\pi\sum_{i}{\alpha_i q(t,iT)}
\end{equation}

in which \textit{h} is the modulation index, \textit{T} is the signalling interval, \textit{$\alpha_i$} is the information data, and \textit{q(t)} is the phase response of the system. $q(t)$ is related to the frequency response \textit{f(t)} through the following formula:

\begin{equation}
\label{eq:q}
  q(t)=\int_{-\infty}^{t}{f(\tau) d\tau}
\end{equation}

the pulse $f(t)$
in (3) is time-limited in the interval $(0, LT)$, where $L$ is the phase pulse duration expressed in symbol intervals, and must satisfy these conditions:

\begin{equation}
  f(t)=f(LT-t)
\end{equation}

\begin{equation}
  \int_{0}^{LT}{f(\tau) d\tau}=q(LT)= \frac{1}{2}
\end{equation}

In the following, the usual way to implement the nonlinear signal (\ref{eq:s}) through a linear modulation will be described.
\FloatBarrier
\begin{figure} [htp]
\begin{center}
    \includegraphics[width=1\linewidth]{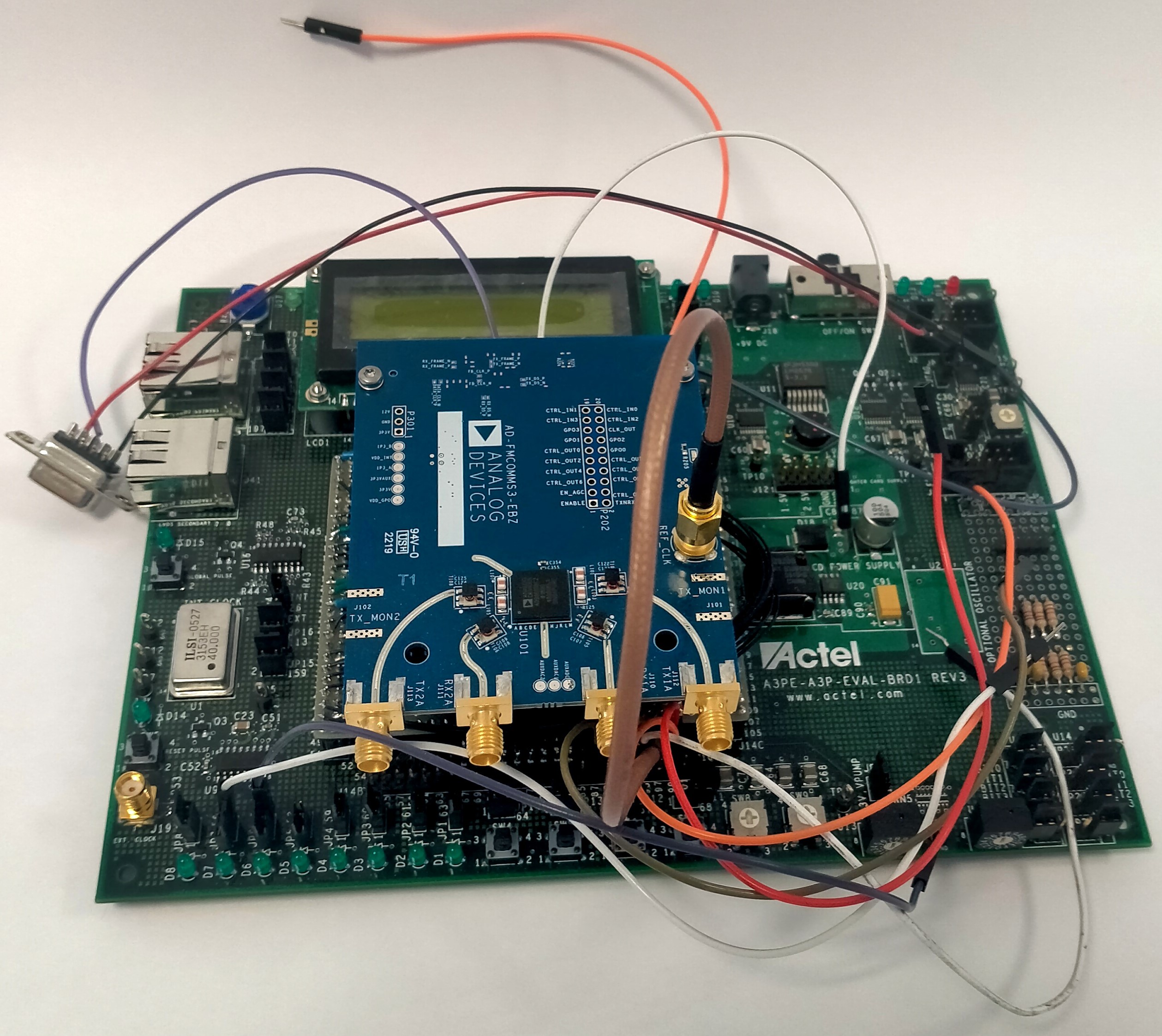}
	\caption{The evaluation board}
	\label{fig:Evaluation board}
\end{center}
\end{figure}
 \subsection{Laurent Decomposition}

In~\cite{artlaur} it was shown that for binary modulation the right-hand side of (1) can be expressed as a superposition of PAM waveforms, while in \cite{pamdec}, this approach was extended to $M$-ary symbols

\begin{equation}
\label{eq:Laurent_mod_sig}
  s(t,{\alpha})= \sum_{k=0}^{Q-1}{\sum_{n}{b_{k,n}c_k(t-nT)}}
\end{equation}

where $Q=2^{L-1}$ and $c_k(t)$ is given by

\begin{equation}
\label{eq:ck}
 c_k(t)= \prod_{i=0}^{L-1}{u(t+iT+\beta_{k,i}LT}), \ \ \ \textrm{$0 \leq k \leq Q-1$}
\end{equation}

The function $u(t)$ is defined as

\begin{equation}
\label{eq:u}
u(t) = \left\{ \begin{array}{lcl}
\sin[2h\pi q(t)]/\sin(h\pi), & 0 \leq t \leq LT \\
\sin[h\pi-2h\pi q(t)]/\sin(h\pi), & LT \leq t \leq  2LT\\
0, & \text{elsewhere}
\end{array}\right.
\end{equation}

The parameter $\beta_{k,i}$ takes the value 0 or 1. In particular $\beta_{k,0}$ is always zero, whereas for any $i$ in the interval $1 \leq i \leq L-1$, $\beta_{k,i}$ is the $i$-th bit in the radix-2 representation of $k$:

\begin{equation}
 k= \sum_{i=1}^{L-1}{2^{i-1}\beta_{k,i}} \ \ \ \ \ \ 0 \leq k \leq Q-1
\end{equation}

The symbols at the input of the linear filters $c_k(t)$ in (\ref{eq:Laurent_mod_sig}), i.e. $b_{k,n}$, are defined as pseudo-symbols and can be computed from the true input symbols $\alpha_i$ as in eq. (10):

\begin{equation}
\label{eq:ps}
 b_{k,n}= exp\left\{jh\pi\left[\sum_{m=-\infty}^{n}{\alpha_m} - \sum_{i=0}^{L-1}{\alpha_{n-i}\beta_{k,i}}\right]\right\}
\end{equation}
\\
\subsection{Space telemetry transmitter}
The PCM/FM modulator described above will be incorporated into the evaluation board shown in Fig. \ref{fig:Evaluation board}. The following Fig. \ref{fig:TX external case} represents a three-dimensional rendering of the final transmitter (TX) case designed to host the hardware realization.
\FloatBarrier
\begin{figure} [htp]
\begin{center}
    \includegraphics[width=0.9\linewidth]{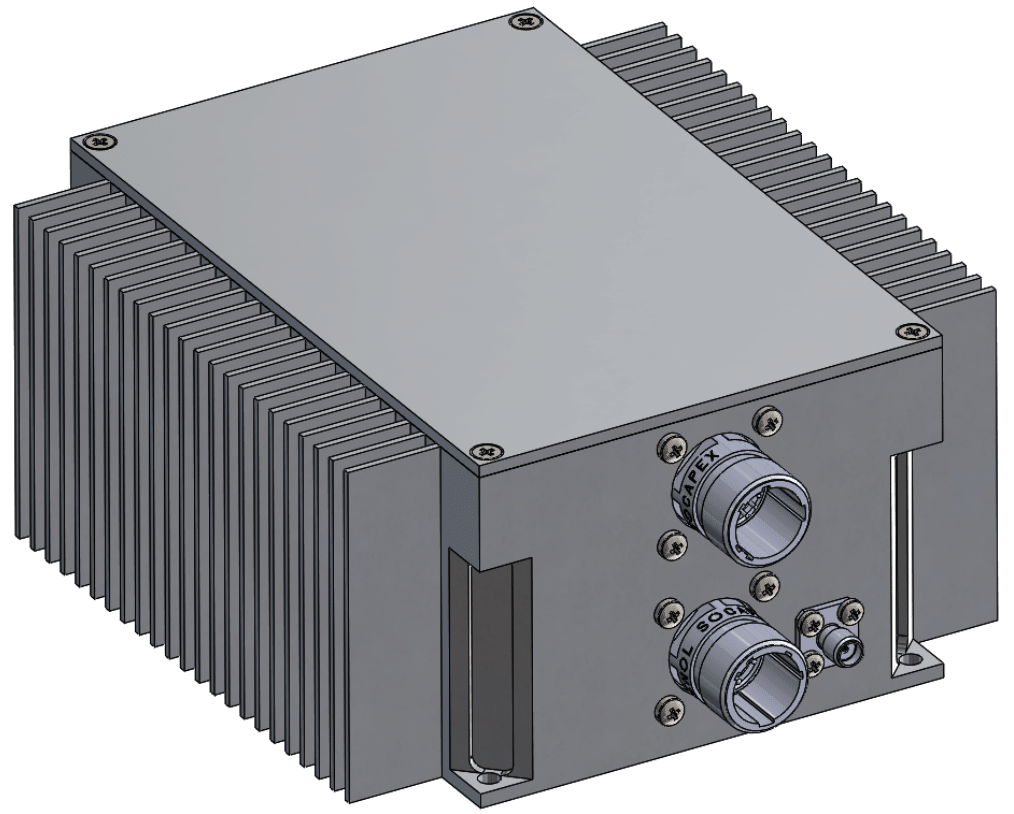}
	\caption{A 3D rendering of TX external case}
	\label{fig:TX external case}
\end{center}
\end{figure}

\section{Proposed PCM/FM transmitter implementation}
\subsection{PCM/FM implementation}
\label{PCM/FM implementation}
A transmitter was initially realised in MATLAB using the telemetry-standard PCM/FM modulation to compare performance with the approximated implementations, that is the one with the first Laurent filter and the one with the MMSE filter. 

In this preliminary work, we focused on this modulation to prove the feasibility of the approach. The presented results can be easily extended to other partial response modulations to be used in telemetry applications. 

The main system parameters were set as follows:

\begin{itemize}

\item $L = 2$
\item $h = 0.7$
\item $\alpha_i = \pm1$ (antipodal bits)
\item $T=1$ (normalized symbol time)
\item $T_c = 1/8$ (sampling time)
\end{itemize}

As a first step, $10000$ antipodal bits $\alpha_i$ were generated randomly to be modulated with an oversampling factor equal to $T/T_C$.

PCM/FM would require a $6^{th}$-order Bessel filter, but a Raised Cosine (RC) filter is an acceptable approximation, more convenient for digital implementation.
The RC filter used to approximate a $6^{th}$-order Bessel filter is described by the following impulse response:

\begin{equation}
f(t) = \left\{ \begin{array}{lcl}
\frac{1}{2LT}\left[1-cos(\frac{2\pi t}{LT})\right], & 0 \leq t \leq LT \\
\\0, & \text{elsewhere}
\end{array}\right.
\end{equation}
The RC filter performance is comparable to the one achieved with Bessel, or even better~\cite{bessrrc}. 






\subsection {Laurent decomposition implementation}
In order to use equation (\ref{eq:Laurent_mod_sig}) to generate the complex envelope of the PCM-FM modulated signal, the filter impulse response $c_k(t)$ is computed for $L=2$, according to the parameters defined in section \ref{PCM/FM implementation}, and using equations (\ref{eq:ck}-\ref{eq:u}).

The pseudo-symbols $b_{k,n}$ are computed using (\ref{eq:ps}), where for $L=2$:
\begin{equation}
\beta_{k,i}= \left( \begin{array}{cc} 0 & 0 \\
0 & 1 \\\end{array} \right)
\end{equation}

\subsection{One component approximation with enhancement performance}
In order to improve the performance achievable with only one linear component representation, a new filter
\begin{equation}
    \label{def_c_w}
    \mathbf{c}_w=[c_w(0),c_w(1),...,c_w(N-1)]
\end{equation}
has been obtained by solving the following Wiener-Hopf equation \cite{haykin2002adaptive}
\begin{equation}
\label{WHeq}
\mathbf{c}_w= \mathbf{R}^{-1}_{b_{0}}\mathbf{r}_{s(t,\alpha) b_{0,n}} \\
\end{equation}
Where $\mathbf{R}_{b_{0}}$ is the autocorrelation matrix of the pseudosymbols $\{b_{0,n}\}$, and $\mathbf{r}_{s(t,\alpha) b_{k,n}}$ is the correlation between the modulated signal $s(nT_c,\alpha)$ and the pseudo-symbols  $\{b_{0,n}\}$. The values assigned to the above autocorrelation  and cross-correlation matrices have been estimated by averaging on more than $8\cdot 10^5$ samples.

Eq. \eqref{WHeq} allows minimizing the MSE between the complete modulated signal representation and the signal representation obtained by filtering the pseudo-simbols $\{b_{0,n}\}$ of (\ref{eq:Laurent_mod_sig}) with $\mathbf{c}_w$.

\section{Simulation and experimental results}
\subsection{Matlab performance evaluation}


A preliminary performance evaluation of the proposed approximated solution involved comparison between the modulated signal by using $\mathbf{c}_w$ with the one obtained by the first Laurent filter $\mathbf{c}_0$. 
From Table I it is possible to appreciate a gain of 3.6 dB with respect to the single-component Laurent approximation.


\begin{table}[ht]
    \caption{Matlab MSE comparison}
    \centering
    \begin{tabular}{ccc} 
    \hline
         \textbf{Tx filter} & \textbf{Oversampled signal} &\textbf{sampled at $1/T$} \\
        \hline
        \\
       $\mathbf{c}_0+\mathbf{c}_1$  & -283.4dB & -292.3dB \\
$\mathbf{c}_0$  & -29.4dB & -57.2dB \\
$\mathbf{c}_w$ & -33.0dB & -60.8dB \\\\
        \hline
    \end{tabular}\label{table:MSE comparison}
\end{table}

The very low value of the MSE related to the comparison between the modulated signal generated by (\ref{eq:s}) and the one built following (\ref{eq:Laurent_mod_sig}) corresponds to the floating point Matlab precision, since the complete Laurent decomposition is an exact representation of the original nonlinear signal.



\subsection{FPGA performance evaluation}
Due to the applicative aim of this work, a field programmable gate array (FPGA) design has been implemented and tested through a hardware description language (HDL)-based simulation that took into account some hardware-related implementation aspects: 16 bits fixed-point operations on signals quantized with 12 bits at the input of the AD9361 digital-to-analog quadrature upconverter.

\begin{figure} [htp]
\begin{center}
    \includegraphics[width=1\linewidth]{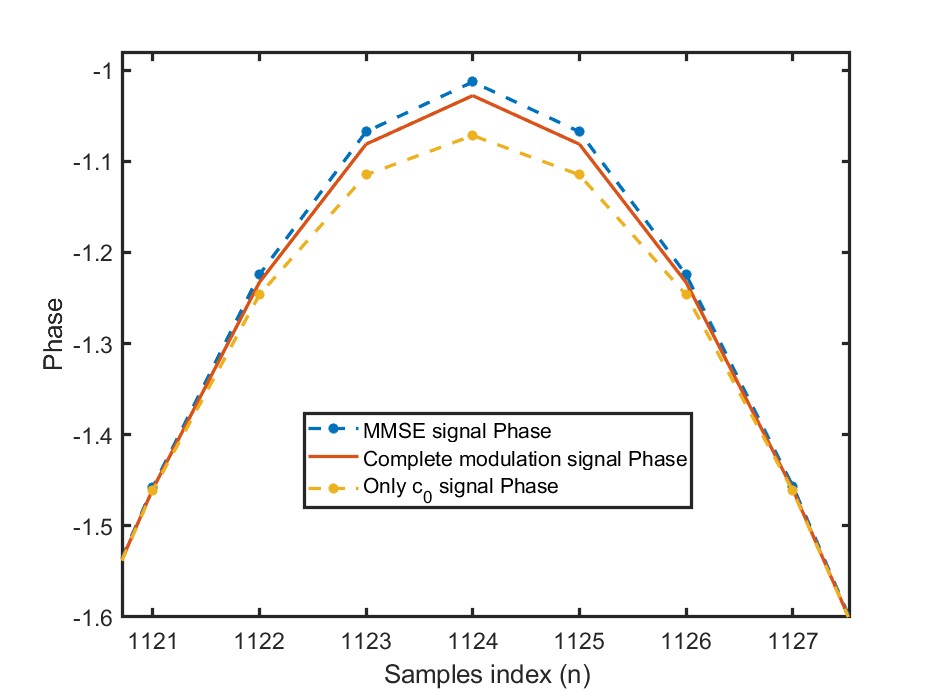}
	\caption{Phase comparison detail}
	\label{fig:Phase comparison detail}
\end{center}
\end{figure}

The FPGA runs at 40 Msps, hence comparison of results with MATLAB output at an oversampling rate of 8 required setting the output FPGA data rate at 5 Mbps.
In order to evaluate how the proposed one-filter modulated signal is closer to the original nonlinear one, a comparison among the MMSE signal phase and the ones generated with full Laurent decomposition and approximate ($\mathbf{c}_0$ only) reconstruction is done in Fig. \ref{fig:Phase comparison detail} by representing the respective phase trends across a number of samples. The sample set was selected randomly among those where the comparison was deemed most significant.

As it can be appreciated, the MMSE approximated phase is closer to the complete nonlinear one with respect to the $\mathbf{c}_0$ only Laurent approximation.

\begin{figure} [htp]
\begin{center}
    \includegraphics[width=1\linewidth]{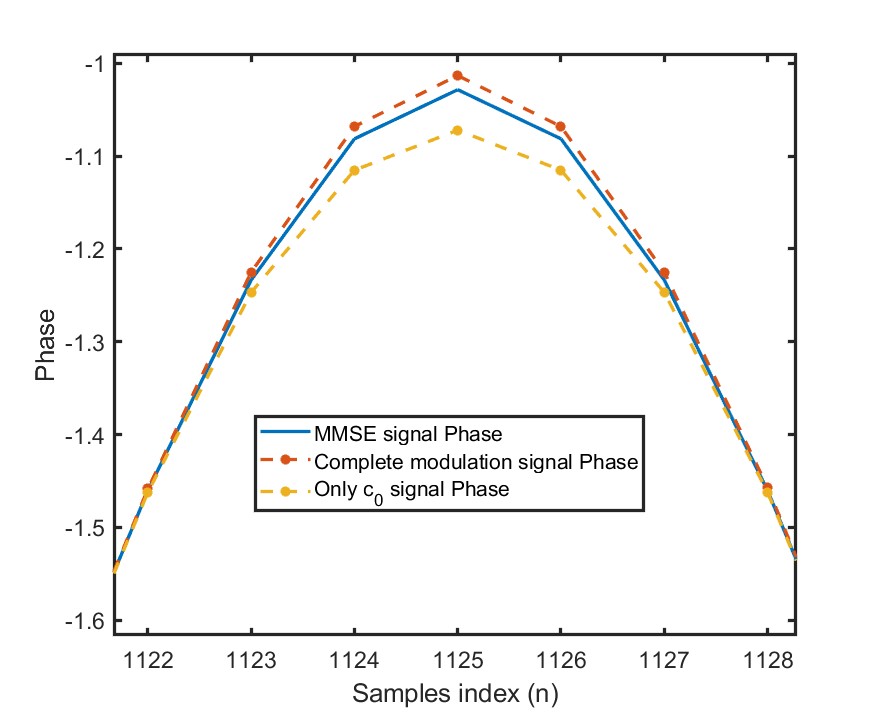}
	\caption{FPGA: phase comparison detail}
	\label{fig:zoomed FPGA phase}
\end{center}
\end{figure}

Fig. \ref{fig:zoomed FPGA phase} shows instead a similar comparison among signals reconstructed in the simulated FPGA realization.

A numerical comparison is provided in Table \ref{table:FPGA MSE comparison} where MSE values are reported for the different approximate representations. For sake of compactness, results are reported only for the oversampled signal.

\begin{table}[ht]
    \caption{FPGA MSE comparison}
    \centering
    \begin{tabular}{cc} 
    \hline
          \textbf{Tx filter} &\textbf{Oversampled signal} \\
        \hline
        \\
$\mathbf{c}_0+\mathbf{c}_1$  & -67.9dB \\
$\mathbf{c}_0$  & -29.7dB \\
$\mathbf{c}_w$ & -33.4dB \\\\
        \hline
    \end{tabular}\label{table:FPGA MSE comparison}
\end{table}

The performance gain provided by the MMSE $\mathbf{c}_w$ filter with respect to using only the first component $\mathbf{c}_0$ of the Laurent representation equals 3.7 dB. The 0.1 dB difference with respect to the floating-point evaluation -based on Matlab- is due to the fixed-point representation used in HDL simulations: 12 bits for signals and 16 bits for internal operations.
\section{conclusions}
In this paper, a novel linear architecture for space telemetry transmitters has been presented. The proposed architecture ensures good performance in terms of spectral shape, while offering two significant advantages:
\begin{enumerate}
    \item a reduced amount of hardware components required for reconstructing the signal;
    \item additional flexibility in terms of different modulations that can be implemented with the same space-qualified hardware.
\end{enumerate}
After simulation and test in the FPGA environment, the proposed solution has reported a 3.7 dB MSE gain over the modulated signal with respect to the first component of Laurent decomposition. It is to be noted that this gain was achieved without increasing the complexity of the system. On the contrary, this latter was reduced thanks to the use of a single signal-reconstructing filter at the transmitter end. 

In a context where both hardware and energy are becoming increasingly scarcer and costlier, another great advantage of this transmitter is its SDR-like nature: it allows using the same space-qualified HW for different purposes, by simply reprogramming it. Upgrades can also be implemented, including implementation of more efficient modulations like SO-QPSK and linear M-QAM, through SW modifications.

This is a factor in limiting space debris, as it allows the device to maintain the required data rates even as it ages, pushing replacement further in the future. 
Programmes for future work involve application of the proposed decomposition to the first two components of the Perrin Rice decomposition for SO-QPSK modulation. A test with the real transmitter will be performed at different data rates to assess how the technique performs in a real-world environment.  

Ongoing works are devoted to test the presented FPGA design in the evaluation board shown in Fig. \ref{fig:Evaluation board}, while the final design will consider the 3D rendering of Fig. \ref{fig:TX external case}.

An interesting future development could extend the proposed transmitter to a distributed multiple-input, multiple output (MIMO) approach like the one envisaged in \cite{WISEE-DMIMO}.

\ifCLASSOPTIONcaptionsoff
  \newpage
\fi



%

%
\label{Bibliography}
\bibliographystyle{IEEEtran}  
\bibliography{Bibliography} 

\begin{thebibliography}{10}
\providecommand{\url}[1]{#1}
\csname url@samestyle\endcsname
\providecommand{\newblock}{\relax}
\providecommand{\bibinfo}[2]{#2}
\providecommand{\BIBentrySTDinterwordspacing}{\spaceskip=0pt\relax}
\providecommand{\BIBentryALTinterwordstretchfactor}{4}
\providecommand{\BIBentryALTinterwordspacing}{\spaceskip=\fontdimen2\font plus
\BIBentryALTinterwordstretchfactor\fontdimen3\font minus
  \fontdimen4\font\relax}
\providecommand{\BIBforeignlanguage}[2]{{%
\expandafter\ifx\csname l@#1\endcsname\relax
\typeout{** WARNING: IEEEtran.bst: No hyphenation pattern has been}%
\typeout{** loaded for the language `#1'. Using the pattern for}%
\typeout{** the default language instead.}%
\else
\language=\csname l@#1\endcsname
\fi
#2}}
\providecommand{\BIBdecl}{\relax}
\BIBdecl

\bibitem{9581544}
C.~Ramu, A.~K. Kushwaha, and P.~Tandon, ``Analysis of aerospace vehicle pcm
  telemetry link in various indoor environments,'' in \emph{2021 2nd
  International Conference on Range Technology (ICORT)}, 2021, pp. 1--6.

\bibitem{9044986}
H.~Xiao, X.~Liu, and L.~Zhou, ``Design of pcm/fm baseband modulation
  transmitter based on software radio,'' in \emph{2018 Eighth International
  Conference on Instrumentation \& Measurement, Computer, Communication and
  Control (IMCCC)}, 2018, pp. 714--717.

\bibitem{9530180}
R.~M. Nigam and P.~M. Pradhan, ``Development of improved soqpsk based data
  transmission over aeronautical telemetry link,'' in \emph{2021 National
  Conference on Communications (NCC)}, 2021, pp. 1--6.

\bibitem{artlaur}
P.~Laurent, ``Exact and approximate construction of digital phase modulations
  by superposition of amplitude modulated pulses (amp),'' \emph{IEEE
  Transactions on Communications}, vol.~34, no.~2, pp. 150--160, 1986.

\bibitem{pamdec}
U.~Mengali and M.~Morelli, ``Decomposition of m-ary cpm signals into pam
  waveforms,'' \emph{IEEE Transactions on Information Theory}, vol.~41, no.~5,
  pp. 1265--1275, 1995.

\bibitem{zhipeng_h}
Y.~F. Zhipeng~Xi, Jiang~Zhu, ``Low-complexity detection of binary cpm with
  small modulation index,'' \emph{ieee communications letters, vol. 20, no. 1},
  pp. 57--60, 2016.

\bibitem{zhramichanest}
A.~S. Rami~Othman, Yves~Lou, ``Joint channel estimation and detection of soqpsk
  using the pam decomposition,'' \emph{2018 25th International Conference on
  Telecommunications (ICT)}, pp. 1--5, 2018.

\bibitem{jsac98}
P.~Savazzi, L.~Favalli, E.~Costamagna, and A.~Mecocci, ``A suboptimal approach
  to channel equalization based on the nearest neighbor rule,'' \emph{IEEE
  Journal on Selected Areas in Communications}, vol.~16, no.~9, pp. 1640--1648,
  1998.

\bibitem{PerrinRice}
E.~Perrins and M.~Rice, ``Pam decomposition of m-ary multi-h cpm,'' \emph{IEEE
  Transactions on Communications}, vol.~53, no.~12, pp. 2065--2075, 2005.

\bibitem{LamoralCoines2021CCSDS1T}
A.~L. Coines and V.~P.~G. Jim{\'e}nez, ``Ccsds 131.2-b-1 transmitter design on
  fpga with adaptive coding and modulation schemes for satellite
  communications,'' \emph{Electronics}, 2021.

\bibitem{bessrrc}
R.~A. Peters, ``How to make a digital satellite link simulation to compare rrc
  and bessel uplink filters,'' \emph{AIAA International Communications
  Satellite Systems Conference}, pp. 1--14, 2006.

\bibitem{haykin2002adaptive}
S.~Haykin, \emph{Adaptive Filter Theory}.\hskip 1em plus 0.5em minus
  0.4em\relax Pearson Education, 2002.

\bibitem{WISEE-DMIMO}
P.~Savazzi and A.~Vizziello, ``Carrier synchronization in distributed mimo
  satellite links,'' in \emph{2015 IEEE International Conference on Wireless
  for Space and Extreme Environments (WiSEE)}, 2015, pp. 1--6.

\end{thebibliography}






\end{document}